\documentclass[prd,superscriptaddress,nofootinbib,eqsecnum,notitlepage,twocolumn,amsmath,amssymb,floatfix]{revtex4} 

\usepackage[utf8]{inputenc}
\usepackage[dvips]{graphicx}
\usepackage{bookmark}
\usepackage{dcolumn}
\usepackage{bm}
\usepackage{color}
\usepackage{ulem} 
\usepackage{url}
\usepackage{amsmath,amssymb}
\usepackage{slashed}
\usepackage{float}
\usepackage{tikz}
\usepackage{setspace}
\usepackage{wasysym}

\usetikzlibrary{shapes,arrows,positioning,automata,backgrounds,calc,er,patterns}
\usepackage[compat=1.1.0]{tikz-feynman}
\usepackage{amssymb}   
\usepackage{amsmath}   
\usepackage{graphicx}   
\usepackage{subfigure}
\usepackage{hyperref}
\usepackage{float}
 
\begin{document}

\title{Chiral symmetry breaking in the pseudo-quantum electrodynamics with non-Abelian four-fermion interactions}
\author{Qiao Yang}
\affiliation{Beijing National Laboratory for Condensed Matter Physics, Institute of Physics, Chinese Academy of Sciences, Beijing 100190, China}
\affiliation{University of Chinese Academy of Sciences, Beijing 100049, China}
\author{Yu-Biao Wu}
\affiliation{Beijing National Laboratory for Condensed Matter Physics, Institute of Physics, Chinese Academy of Sciences, Beijing 100190, China}

\author{Wu-Ming Liu}  \email{wliu@iphy.ac.cn}
\affiliation{Beijing National Laboratory for Condensed Matter Physics, Institute of Physics, Chinese Academy of Sciences, Beijing 100190, China}
\affiliation{University of Chinese Academy of Sciences, Beijing 100049, China}
\affiliation{Songshan Lake Materials Laboratory, Dongguan, Guangdong 523808, China}

\date{\today}
\begin{abstract}
	In the context of 2+1 dimensional Dirac materials, we consider electromagnetic interactions alongside a type of spin-dependent Hubbard interaction. The former is described by PQED theory, while the latter corresponds to an effective theory represented by the $SU(N_c)$ Thirring model. Employing Hubbard-Stratonovich transformation and large N expansion in the model yields a non-local $SU(N_c)$ Yang-Mills action. Subsequently, we solve Schwinger-Dyson equations to obtain the self-energy function of the fermion propagator, from which we determine the critical fermion flavor number $N_f ^c$ and critical fine structure constant $\alpha_c$ indicative of chiral symmetry breaking. Our findings suggest that as the non-Abelian color number $N_c$ increases, the minimum value of the critical fermion flavor number monotonically increases, while the maximum value of the critical fine structure constant decreases accordingly, rendering the system more susceptible to chiral symmetry breaking.
\end{abstract}
	\maketitle
	
	\section{\textbf{Introduction}}

	1+2 dimensional quantum field theories is a proper framework for describing  electron interactions in thin material, for instance, silicene, graphene, and transition metal dichalcogenides \cite{neto2009electronic}.In particular, the emergence of both massive and massless dirac excitations in two-dimensional model, such as tight-binding fermions in a honeycomb lattice configuration or haldane model, has built a connection between high-energy and condensed matter physics.
	
	In these 1+2 quantum field theory models, pseudo quantum electrodynamics (PQED)\cite{marino1993quantum,magalhaes2020pseudo,fernandez2020renormalization,fernandez2021influence} has attracted some attentions. It's a non-local model obtained through dimensional reduction, ensuring Coulomb interactions among electrons in two-dimensional materials. This places fermions in the plane while the gauge fields mediating Coulomb interactions between fermions reside in 3+1-dimensional spacetime.
    In some literatures, the possibility of chiral symmetry breaking(or dynamical mass generation) in PQED has been discussed\cite{alves2013chiral,nascimento2015chiral,gorbar2001dynamical,cuevas2020chiral,albino2022electron}.
	Morever, the PQED with four fermion interaction (such as Gross-Neveu ) also has been considered\cite{alves2017dynamical,fernandez2021dynamical}. All these interactions are ableian, 
	however, in some plane Dirac models, considering a type of spin-dependent Hubbard  interaction, non-abelian emerges spontaneously\cite{palumbo2014non}.The dynamical symmetry breaking in PQED with non-abelian four-fermion interactions has not been investigated until now. And we expect this non-abelian interaction to be non-trivial because the dynamical behavior of the fermions will depend on the dimensionality of the non-abelian representation $N_c$.
	 
	 In this paper, we describe the dynamical mass generation in PQED coupled to a non-Abelian four fermions interaction, i.e.,$SU(N_c)$ Thirring term, at zero temperature. First, we use a Hubbard-Stratonovich to decouple the $SU(N_c)$ Thirring interaction to obtain a non-abelian auxiliary field. Then the gauge-field propagator and the auxiliary-field propagator are calculated in order of $1/N_f$. Thereafter, we use this result into the Dyson-Schwinger equation for the electro self-energy and calculate the full electron propagator in the dominant order of $1/N_f$ in the nonperturbative limit.
	 Final,we obtain the critical fermion flavor number $N_f ^c$ and critical fine structure constant $\alpha_c$ indicative of chiral symmetry breaking in the system, both of which depend on the non-Abelian color number $N_c$.
	  
	  The outline of this paper is the following. In sec.~II. we show our model within the PQED approach and the $SU(N_c)$ Thirring model.In Sec.~III, we utilized the HS transformation to decouple the non-Abelian four-fermion interaction by introducing a set of auxiliary fields, and derived the corresponding Feynman rules. In Sec.~IV, we calculated the one-loop self-energy of the gauge boson fields and auxiliary fields. In Sec.~V, we solved the fermion self-energy function using the Dyson-Schwinger equation. In Sec.~VI, based on the fermion self-energy function, we derived analytical expressions for the critical number of fermion flavors $N^c_f$ and the critical fine structure constant $\alpha_c$. In Sec.~VII, we summarized our main results and provided a detailed derivation of the effective Hamiltonian for the spin-type Hubbard interaction to the $SU(N_C)$ non-Abelian interaction in the appendix.

	\section{ model}
	\label{MP-PlanarCurent}
	The Euclidean non-Abelian Lagrangian reads 
	\begin{align}\label{action}
		{\cal L}&=\frac{1}{2} \frac{F_{\mu \nu} F^{\mu \nu}}{\sqrt{-\partial^2}}+ i\bar\psi_{I,i}\slashed{\partial}\psi_{I,i}+j^{\mu}\, A_{\mu} \notag,\\
			&-\frac{U}{2}(\bar\psi_{I,i} \gamma^\mu T^a _{ij} \psi_{I,j})(\bar\psi_{I,i} \gamma_\mu T^a _{ij} \psi_{I,j}).
	\end{align}
	where $\psi_{I,i}$  is a four-component Dirac field which is  be confined in a plane. The index $i$ denote the color index means $i=1,...,D(N_c)$ where $D(N_c)$ is the dimension of the representation. The flavor index $I=1,...,N_f$,  and $\bar\psi_{I,i} =\psi_{I,i}^\dagger\gamma^0$ is its adjoint. $j^{\mu}=e\bar\psi_{I,i} \gamma^{\mu}\psi_{I,i}$ is the matter current,  $F^{\mu \nu}$ is the field intensity tensor of the local U(1) Gauge field $A_\mu$ with a dimension of energy(i.e. $[A_\mu]=[M]$). The nonlocal operator $\frac{1}{\sqrt{-\Box}}$  make the $QED_3$ action into the $PQED_{1+2}$ which can reproduce the coulomb interaction rather than logarithmic interaction for two charged in-plane fermions.  $\gamma^\mu$ is a 4-dimensional representation of the cliford algebra and satisfies$\{\gamma^\mu,\gamma^\nu \}=-2 \delta^{\mu \nu}$, $T^a_{ij}$ is the generator matrices of $SU(N_c)$ which obey commutation relations of the form $[T^a,T^b]=i f^{abc}T^c$, we will use $T^a(R)$ to denote the representation of that group is R, and  $D(R)$ is the dimension of the representation(Specifically, the representations considered in this paper are in the fundamental representation), the colors index a runs from 1 to the $N_c ^2-1$ for the $SU(N_c)$ group. 
$U$ is the coupling constant related to the non-abelian four-fermion interaction. The dimension of $U$ is the inverse of energy, namely, $[U]=[M]^{-1}$.
	\section{Hubbard-Stratonovich transformation}
	Let us first start with the non-Abelian four fermionic term in Eq~(\ref{action}), i.e., the non-Abelian $SU(N_c)$ thirring action ${\cal L}_{NAT}$, given by
	\begin{equation}\label{NAT}
		{\cal L}_{NAT}=i\bar\psi_{I,i}\slashed{\partial} \psi_{I,i}-\frac{U}{2}(\bar\psi_{I,i} \gamma^\mu T^a _{ij} \psi_{I,j})(\bar\psi_{I,i} \gamma_\mu T^a _{ij} \psi_{I,j}).
	\end{equation}
	\\
	\\
	The first step is to introduce the $N_f$ parameter into the action, we will redefine a new coupling constant $u=UN_f$ such that $u$ is be fixed for the large $N_f$. Next, we use the Hubbard-Stratonovich transformation to introduce a set of vector field called $a^a _\mu$, and we can obtain a three-linear vertex interaction instead of the original four fermionic vertices. The specific steps are as follows: 
	first, we introduce
	\begin{align}\label{au}
		\mathbf{1}&=\int D a_\mu e^{\int -\frac{1}{2u}tr( a_\mu a^\mu)} \notag \\
	&=\int D a_\mu ^a e^{-\frac{T(R)}{2u}a_\mu ^a a^{\mu a}}\nonumber \notag\\
	&=\int D a_\mu ^a e^{-\frac{1}{2u}a_\mu ^a a^{\mu a}} .
\end{align}
	in the second line, we used $tr(a^\mu a_\mu)=a_\mu ^a a^{\mu b}tr[T^aT^b]=T(R)a_\mu ^a a^{\mu a}$, and we have absorbed $T(R)$ into the integral measure in the third line.
	Next, we have the transformation
	\begin{align}
		\mathcal{L}_{NAT} &\rightarrow \mathcal{L}_{NAT} + \frac{1}{2u}\left(a_\mu^a -\frac{u}{\sqrt{N_f}}(\bar\psi_{I,i} \gamma_\mu T^a_{ij} \psi_{I,j})\right) \nonumber \\
		&\quad \times \left(a^{\mu a}-\frac{u}{\sqrt{N_f}}(\bar\psi_{I,i} \gamma^\mu T^a_{ij} \psi_{I,j})\right).
		\end{align}
	
	which converts the action into
	\begin{align}
		{\cal L}_{NAT}=i\bar\psi_{I,i}\slashed{\partial} \psi_{I,i} + \frac{1}{2u} a_\mu ^a a^{\mu a}-\frac{1}{\sqrt{N_f}}a^{\mu a} \bar\psi_{I,i} \gamma_\mu T^a _{ij} \psi_{I,j}.
    \end{align}
	Finally, combine with the PQED Lagrangian of the large $N_f$ expansion, Eq.~(\ref{action}) becomes
	\begin{align}\label{ssunpqed}
    	{\cal L}&=\frac{1}{4} F_{\mu \nu}\left[\frac{2}{\sqrt{-\partial^2}}\right] F^{\mu\nu} +i\bar\psi_{I,i}\slashed{\partial} \psi_{I,i}+e\bar\psi_{I,i} \gamma_\mu \psi_{I,i}A_{\mu} \notag \\
		&+\frac{1}{2u} a_\mu ^a a^{\mu a}-\frac{1}{\sqrt{N_f}}a^{\mu a} \bar\psi_{I,i} \gamma_\mu T^a _{ij} \psi_{I,j}.
\end{align}
	Therefore, the feynman rules we can read from the lagrangian are as follows:
	\\
	1.the bare gauge-field propagator, in the Landau gauge, is\\
	\begin{equation}
	\begin{tikzpicture}
	\begin{feynman}
	\vertex (a);
	\vertex [right=1cm of a, label=\(\mu\)](b);
    \vertex [right=2cm of b, label=\(\nu\)](c);
	\diagram*
	{
	(b) -- [green,charged boson] (c),
	};
	\end{feynman}
	\end{tikzpicture}
	=G_{0,\mu\nu}=\frac{P_{\mu\nu}}{2\sqrt{p^2}}	
		\end{equation}

2.the bare auxiliary vector--field propagator is\\
	\begin{equation}
	\begin{tikzpicture}
	\begin{feynman}
	\vertex (a);
	\vertex [right=1cm of a, label=\(a\mu\)](b);
    \vertex [right=2cm of b, label=\(b\nu\)](c);
	\diagram*
	{
	(b) -- [red,charged scalar] (c),
	};
	\end{feynman}
	\end{tikzpicture}
	=\Delta^{\mu\nu}_{0,ab}=\frac{\delta_{ab}\delta^{\mu\nu}}{1/\mu}
		\end{equation}

3.the bare fermion propagator is\\
\begin{equation}
	\begin{tikzpicture}
	\begin{feynman}
	\vertex (a);
	\vertex [right=1cm of a, label=\(i\)](b);
    \vertex [right=2cm of b, label=\(j\)](c);
	\diagram*
	{
	
	(b) -- [fermion] (c),
	};
	\end{feynman}
	\end{tikzpicture}
	=S^0 _{f,ij}=\frac{\delta_{ij}}{\gamma^\mu p_\mu}
		\end{equation}
		
4.the fermion-auxiliary field vertex is\\
	\begin{equation}
	\begin{tikzpicture}
	\begin{feynman}
	\vertex (b);
	\vertex [above=1cm of a,label=\(a\,\mu\)] (a);
	\vertex [left=1cm of b,label=\(i\)] (c);
	\vertex [right=1cm of b,label=\(j\)] (d);
	\diagram*
	{
	(a)--[red,charged scalar](b),
	(b) -- [anti fermion] (c),
	(b) -- [fermion](d),
	};
	\end{feynman}
	=sfsfsfwf
	\end{tikzpicture}
	=-\frac{1}{\sqrt{N_f}}\gamma^\mu T^a_{ij} 
		\end{equation}
		
5.the fermion-gauge filed vertex is\\
\begin{equation}
	\begin{tikzpicture}
	\begin{feynman}
	\vertex (b);
	\vertex [above=0.2cm of a,label=\(\mu\)] (a);
	\vertex [left=1.3cm of b,label=\(i\)] (c);
	\vertex [right=1.3cm of b,label=\(j\)] (d);
	\diagram*
	{
	(a)--[green,charged boson](b),
	(b) -- [anti fermion] (c),
	(b) -- [fermion](d),
	};
	\end{feynman}
	=sfsfsfwf
	\end{tikzpicture}
	=-e\gamma^\mu \delta_{ij}
		\end{equation}

	Next, we would like to calculate the possibility of dynamical mass generation due to interactions. The natural method to investigate such phenomena is the Schwinger-Dyson equation for the electron propagator, which yields nontrivial results at large coupling constants. This is a system of coupled integral equations, connecting all of the full propagators in the theory. To obtain an analytical solution for the electron, we must approximate both the gauge and scalar propagators. The easiest approach would be to consider only the bare propagators. Here, nevertheless, we choose to apply the large-$N_f$ expansion, which allows us to include quantum fluctuations in these propagators.
Before, we use the Schwinger-Dyson equation for the electron propagator, we need to obtain the propagators for the Gauge and scalar field. In the next section, we apply the large-$N_f$ expansion in order to obtain these full propagators.

	\section{Exact propagator for both $a_\mu ^a $ and $A_\mu$ in the Large-$N_f$ expansion}
In this section, we apply the large-$N_f$ expansion  to obtain the full propagators of the non-abelian auxiliary fields and the gauge fields. The Schwinger-Dyson equation for the auxiliary vector field $a^a_\mu$ reads
\begin{equation}
	\Delta_{\mu\nu} ^{-1,ab}(p)=\Delta_{0,\mu\nu}^{-1,ab}(p)-\Pi^{ab}_{\mu\nu}(p)\,\label{self of vector}
\end{equation}
where the $\Delta_{\mu\nu} ^{-1,ab}(k)$ is the full propagator. $\Pi^{ab}_{\mu\nu}(k)$, the
self-energy correction of the vector, is the quantum correction due to interaction withe matter field. This is given by the fermionic loop, namely,
\begin{eqnarray}
\Pi^{ab}_{\mu\nu}(p)= \nonumber \\
\begin{tikzpicture}
	\begin{feynman}
	\vertex (m);
	\vertex [label=\(a \mu \),right=2cm of m](a);
	\vertex [right=1.2cm of a] (b);
	\vertex [right=1.3cm of b] (c);
	\vertex [label=\(b \nu \),right=1.2cm of c] (d);
	\diagram*
	{(a)-- [red,charged scalar,edge label'=\(p\)] (b),
	(b)-- [anti fermion,half left,edge label=\(k\)] (c),
	(c)-- [anti fermion,half left,edge label=\(p+k\)](b),
	(c) -- [red,charged scalar,edge label'=\(p\)] (d),
};
	\end{feynman}
	\end{tikzpicture}
	\end{eqnarray}
Therefore, according to the feynman rules, we have:
\begin{eqnarray}
	\Pi^{ab}_{\mu\nu}(p)&=&(-1)Tr \int \frac{d^3 k}{(2\pi)^3}\gamma_\mu T^a _{ij}S_{f,jl}(k)\gamma_\nu T^b_{lm}S_{f,mi}(p+k) \nonumber \\
	&=&(-1)tr(T^aT^b)Tr \int \frac{d^3 k}{(2\pi)^3}\gamma_\mu S_{f}(k)\gamma_\nu S_{f}(p+k) \nonumber \\
	&=&(-1)T(R)\delta^{ab}Tr\int \frac{d^3 k}{(2\pi)^3}\gamma_\mu S_{f}(k)\gamma_\nu S_{f}(p+k) \nonumber \\
\end{eqnarray}
where we have use $tr(T^aT^b)=T(R)\delta^{ab}$, the R denote the fermions are in a representation R.
After some calculations, we obtain
\begin{equation}
	\Pi^{ab}_{\mu\nu}(p)=-\frac{T(R)}{8}\sqrt{p^2}P_{\mu\nu} \delta^{ab} \label{vector se}
\end{equation}
where the $P_{\mu\nu}=\delta_{\mu\nu}-\frac{p_{\mu\nu}}{p^2}$ is the projection matrix. 

Now,we will
 calculate the full propagator $\Delta_{\mu\nu} ^{ab}(p)$. For the first of all, combine Eq.~(\ref{self of vector}) and Eq.~(\ref{vector se}),we can get the 
 \begin{equation}
 	\Delta_{\mu\nu} ^{-1ab}(p)=(\frac{1}{u}\delta_{\mu \nu} + \frac{T(R)}{8}\sqrt{p^2}P_{\mu\nu})\delta^{ab}
 \end{equation}
  For simplify, we set $\frac{T(R)}{8}=x_0$, and due to $\Delta_{\mu\nu} ^{-1}\Delta_{\nu\rho} (p)=\delta_{\mu\rho}$, we can obtain:
  \begin{equation}
  	\Delta_{\mu\nu} ^{ab}(p)=(u \delta_{\mu \nu} - \frac{\mu x_0 \sqrt{p^2}}{1/u+x_0 \sqrt{p^2}}P_{\mu \nu}) \delta^{ab} \label{p vector}
  \end{equation}
  
  Next, we apply the very same set of approximations for the gauge field. The Schwinger-Dyson equation is
  \begin{equation}
  	G^{-1}_{\mu\nu}=G^{0,-1}_{\mu\nu}-\Xi_{\mu\nu}
  \end{equation}
where the $\Xi_{\mu\nu}$ is the vacuum polarization tensor, and its feynman diagram is:

\begin{tikzpicture}
	\begin{feynman}
	\vertex (m);
	\vertex [label=\(\mu \),right=2cm of m](a);
	\vertex [right=1.2cm of a] (b);
	\vertex [right=1.3cm of b] (c);
	\vertex [label=\(\nu \),right=1.2cm of c] (d);
	\diagram*
	{(a)-- [green,charged boson,edge label'=\(p\)] (b),
	(b)-- [anti fermion,half left,edge label=\(k\)] (c),
	(c)-- [anti fermion,half left,edge label=\(p+k\)](b),
	(c) -- [green,charged boson,edge label'=\(p\)] (d),
};
	\end{feynman}
	\end{tikzpicture}
	
Therefore, we have
\begin{eqnarray}
\Xi_{\mu\nu}(p) &=&(-1)D(R)N_fe^2Tr \int \frac{d^3 k}{(2\pi)^3}\gamma_\mu S_{f}(k)\gamma_\nu S_{f}(p+k) \nonumber\\
&=&-\frac{e^2D(R)N_f}{8}\sqrt{p^2}P_{\mu\nu}
\end{eqnarray}
For summing the self-energies in the large-$N_f$ expansion for PQED, we shall replace $e^2\rightarrow \lambda/N_f$, where $\lambda$ is taken fixed at large $N_f$. This allow us to sum over all of the diagrammatic contributions in order of $1/N_f$, which is an infinite sum, unlike the standard perturbation in $e$. Therefore, the PQED vertex reads $\sqrt{\lambda/N_f} \gamma_\mu$, describing the electromagnetic interaction.
Where the $D(R)$ is the dimension of the representation of the fermion field, and So the full propagator of the gauge field is
\begin{equation}
	G_{\mu\nu}=\frac{P_{\mu\nu}}{\sqrt{p^2}(2+\frac{D(R)\lambda}{8})} \label{p gauge}
\end{equation}

Next, we use the propagators in Eq.~(\ref{p vector}) and Eq.~(\ref{p gauge}) to calculate the full electron propagator $S_f(p)$. Thereafter, we shall use this function to obtain the dynamically generated masses.
\section{Self-energy of fermion}
\begin{figure}[H]
	\centering
	\includegraphics[width=0.45\textwidth]{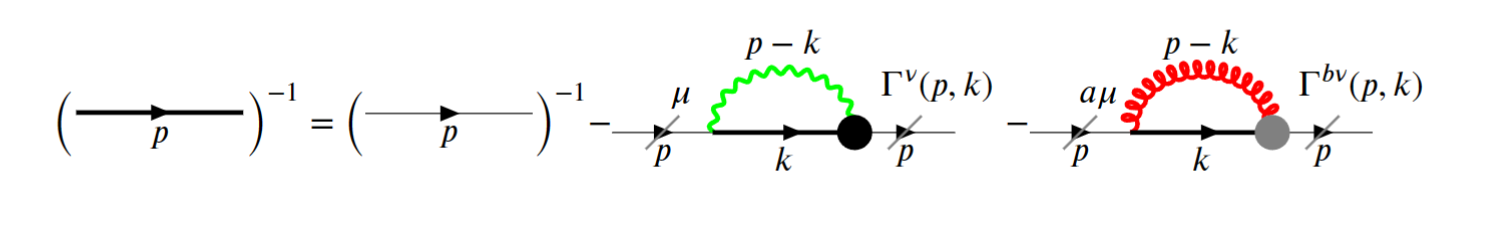}
  \caption{The Schwinger-Dyson equation for the electron.
The left-hand side is the inverse of the full propagator of the
fermion field, the first term in the right-hand side corresponds to
the inverse of the  bare fermion field propagator, and the other terms in the right-hand side are the electron self-energy $\Sigma$. We employ the ladder approximation (also known as the rainbow approximation), which neglects the quantum corrections to the vertex function: $\Gamma^{\nu}(p,k) \rightarrow \gamma^{\nu}$,$\Gamma^{b\nu}(p,k)\rightarrow \gamma^{\nu} T^b$.}
 \label{fermionsself}
\end{figure}

The Schwinger-Dyson equation for the electron reads:
\begin{equation}
S^{-1}_F(p)=S_{0F}^{-1}(p)-\Sigma (p),
\label{sd}
\end{equation}
where $S_{0F}$ and $S_{F}$ are, the bare- and full propagators, respectively, and we ignore the index $S_{F,ij}$ and $\Sigma_{ij}$, for simplify. $\Sigma(p)$ is the electron self-energy, which has two contributions $\Sigma(p)=\Sigma^{\alpha}(p)+\Sigma^u(p)$, where $\Sigma^\alpha(p)$ and $\Sigma^u(p)$ are the electron self-energies due to the electromagnetic and $SU(N_c)$ thirring interactions, respectively. The diagrammatic representation is shown in Fig.~\ref{fermionsself}. The electron-self energies are given by

\begin{eqnarray}\label{sea}
\Sigma^\alpha (p)=\frac{\lambda}{N_f} \int\frac{d^3k}{(2\pi)^3} \gamma^{\mu}S_F(k)\gamma^{\nu}\, G_{\mu\nu}(p-k)\,
\end{eqnarray}
and
\begin{eqnarray}\label{seg}
\Sigma^u (p)=\frac{1}{N_f} \int\frac{d^3k}{(2\pi)^3} \Delta_{\mu\nu}^{ab}(p-k)T^a_{ik} \gamma^\mu S_{F,km}(k)\gamma^\nu T^b_{mj} \nonumber \\
\end{eqnarray}

In order to obtain analytical solutions of the Schwinger-Dyson equation, it is convenient to rewrite the full fermion propagator:
\begin{equation}
S_F^{-1}(p)=p_{\mu}\gamma^{\mu}\,A(p)-\Sigma (p), \label{full1}
\end{equation}
where $A(p)$ is usually called the wavefunction renormalization, and $\Sigma(p)$ the mass function. Inserting Eq.~(\ref{full1}) into Eq.~(\ref{sd}), we obtain the integral equation
\begin{eqnarray}
\label{sigmap}
\Sigma(p)&=&\frac{2\lambda}{N_f}\int\frac{d^3k}{(2\pi)^3}\,\frac{\Sigma(k)\,}{A^{2}(k)k^2+\Sigma^2(k)}\,\frac{1}{\sqrt{(p-k)^2}(2+\frac{D(R)\lambda }{8})} \nonumber \\
&-& \frac{C(R)}{N_f} \int\frac{d^3k}{(2\pi)^3} \frac{2F(u,p,k)\Sigma(k)}{A^{2}(k) k^2+\Sigma^2(k)}  \nonumber \\
&+&\frac{C(R)}{N_f} \int\frac{d^3k}{(2\pi)^3} \frac{3u\Sigma(k)}{A^{2}(k) k^2+\Sigma^2(k)}  
\end{eqnarray}
Where $F(u,p,k)=\frac{ux_0 \sqrt{(p-k)^2}}{\frac{1}{u}+x_0 \sqrt{(p-k)^2}}$.
The expression for \(A(p)\) is approximated as \(1+\mathcal{O}\left(\frac{1}{N_f}\right)\). Therefore, in subsequent calculations, we can approximately treat \(A(p)\) as the constant 1.

\section{\textbf{VI. The Dynamical Solution for $\Sigma(p)$}}

The third term on the rhs of Eq.~(\ref{sigmap}) does not change momentum. Hence, we focus on the dynamical solutions of the mass function driven by a kernel with $p$-dependence. This regime is obtained when both Gauge and auxiliary propagators change momentum with the electron propagator.  
\begin{eqnarray}
\label{sigmap3}
\Sigma(p)&=&\frac{2\lambda}{N_f}\int\frac{d^3k}{(2\pi)^3}\,\frac{\Sigma(k)\,}{k^2+\Sigma^2(k)}\,\frac{1}{\sqrt{(p-k)^2}(2+\frac{D(R)\lambda}{8})} \nonumber \\
&-& \frac{C(R)}{N_f} \int\frac{d^3k}{(2\pi)^3} \frac{2F(u,p,k)\Sigma(k)}{A^{2}(k) k^2+\Sigma^2(k)}  \nonumber \\
\end{eqnarray}

We use spherical coordinates $d^3k=k^2 dk \sin \theta d\theta d\phi$, hence, the polar integral gives a factor $2\pi$. Next, we solve the angular integral in both the first and second terms in the rhs of Eq.~(\ref{sigmap3}). 

Let us first consider only the integral which is proportional do $\lambda$, i.e, the first term of the rhs of Eq.~(\ref{sigmap3}). By defining $u\equiv p^2+k^2-2pk \cos\theta$ and changing the integral variable into $u$, we find
\begin{equation}
\frac{2\lambda}{4 \pi^2 N_f} \frac{1}{(2+\frac{D(R)\lambda}{8})} \int_0^\infty \frac{k^2 dk \Sigma(k)}{k^2+\Sigma^2(k)}\left(\frac{|p+k|-|p-k|}{pk}\right).
\end{equation}
For the second term in the rhs, the same procedure yields
\begin{eqnarray}
&&\frac{2C(R)}{4 \pi^2 N_f p x_0} \int_0^\infty \frac{k dk \Sigma(k)}{k^2+\Sigma^2(k)} \{ [|p+k|-|p-k|\nonumber\\
&-&\frac{1}{x_0 u} \ln \left[\frac{(x_0 u)^{-1}+|p+k|}{(x_0 u)^{-1}+|p-k|}\right]\} \nonumber \\
&-&\frac{2C(R)}{4 \pi^2 N_f} \int_0^\infty \frac{2uk^2 dk \Sigma(k)}{k^2+\Sigma^2(k)}
\end{eqnarray}
Therefore, the integral equation becomes
\begin{eqnarray}
&\Sigma(p)&=\frac{C_2}{p}\int_0^\infty\frac{k dk \Sigma(k)}{k^2+\Sigma^2(k)} \ln \left[\frac{(x_0 u)^{-1}+|p+k|}{(x_0 u)^{-1}+|p-k|}\right]\nonumber\\
&&+\frac{C_1}{p}\int_0^\infty \frac{k dk \Sigma(k)}{k^2+\Sigma^2(k)}\left(|p+k|-|p-k|\right), \label{eqint0}
\end{eqnarray}
where
\begin{equation}
C_1=\frac{2\lambda}{4\pi^2N_f(2+\lambda D(R)/8)}+\frac{2C(R)u}{4\pi^2 N_f (u x_0)},
\end{equation}
and
\begin{equation}
C_2=-\frac{2C(R)u}{N_f (u x_0)^2 4\pi^2}.
\end{equation}

At this level, it is convenient to write a scale-invariant integral equation (without any dimensional parameter). By defining $\Sigma(p)\equiv \sigma(pu)/u$, $x\equiv u p$, $y\equiv u k$, we find: 
\begin{align} \label{eqint1}
	\sigma(x) &= \frac{u C_2}{x}\int_0^\infty\frac{ dy y \sigma(y)}{y^2+\sigma^2(y)} \ln \left[\frac{(x_0)^{-1}+|x+y|}{(x_0)^{-1}+|x-y|}\right] \notag \\
	&\quad +\frac{C_1}{x}\int_0^\infty \frac{dy y \sigma(y)}{y^2+\sigma^2(y)}\left(|x+y|-|x-y|\right).
\end{align}

Using the following approximation:
\begin{equation}
\ln \left(\frac{x_0^{-1}+|x+y|}{x_0^{-1}+|x-y|}\right) \approx  \frac{2 y}{x+x_0^{-1}} \Theta(x-y) 
 +\frac{2 x}{y+x_0^{-1}} \Theta(y-x).
\end{equation}
and introducing the ultraviolet cutoff parameter $\Lambda$, the integral equation (\ref{eqint1}) becomes:
\begin{align}\label{eqint22}
\sigma(x)= & \frac{2 C_1}{x}[\int_0^x \frac{y^2 \sigma(y) dy}{y^2+\sigma^2(y)}+x \int_x^{u \Lambda} \frac{y \sigma(y) dy}{y^2+\sigma^2(y)}] \notag \\
& +u C_2[\int_0^x \frac{y^2 \sigma(y) dy}{y^2+\sigma^2(y)} \frac{2}{x(x+x_0^{-1})}\notag \\
&+\int_x^{u \Lambda} \frac{y \sigma(y) dy}{y^2+\sigma^2(y)} \frac{2}{(y+x_0^{-1})}]. 
\end{align}
From equation (\ref{eqint22}), the first derivative of $\sigma(x)$ can be obtained:
\begin{align}\label{eqint33}
\frac{d \sigma}{d x}&= -\frac{2 C_1}{x^2} \int_0^x \frac{y^2 \sigma(y) dy}{y^2+\sigma^2(y)}\notag \\
&+u C_2 \frac{d}{d x}\left[\frac{2}{x\left(x+x_0^{-1}\right)}\right] \int_0^x \frac{y^2 \sigma(y) dy}{y^2+\sigma^2(y)}.
\end{align}
In particular, we define:
\begin{equation}\label{hs}
h^{-1}(x)=1-\frac{u C_2 x^2}{2 C_1} \frac{d}{d x}\left[\frac{2}{x\left(x+x_0^{-1}\right)}\right].
\end{equation}
Combining equations (\ref{eqint33}) and (\ref{hs}), and considering the high momentum case: $x \gg \sigma(x)$, the integral equation (\ref{eqint1}) can be linearized into a second-order linear differential equation:
\begin{equation} \label{eulereq}
\frac{d}{dx}\left(x^2\frac{d \sigma(x)}{dx}\right)+\frac{N^{c}_f}{4N_f}\sigma(x)=0.
\end{equation}
Its corresponding ultraviolet (UV) and infrared (IR) asymptotic conditions are respectively:
\begin{align}\label{uvirsup}
&\lim_{x\rightarrow \Lambda}\left(x \,\frac{d\sigma(x)}{dx}+\sigma(x)\right)=0,\notag \\
&\lim_{x\rightarrow 0}x^2\,\frac{d\sigma(x)}{dx}=0.
\end{align}
where
\begin{equation}\label{Nc}
N^{c}_f(\lambda)= \frac{2}{\pi^2}\left(\frac{2\lambda}{2+\lambda D(R)/8}+16\frac{C(R)}{T(R)}\right).
\end{equation}
is the critical number of fermion flavors. Noticing that the differential equation (\ref{eulereq}) is an Euler equation, its general solution is:
\begin{equation}\label{EulerSol}
\sigma(x)=A_+ x^{a_{+}}+ A_- x^{a_{-}}.
\end{equation}
where $a_{\pm}=-\frac{1}{2}\pm\frac{1}{2}\sqrt{1-N^c _f/N_f}$, and $A_+$, $A_-$ are arbitrary constants. For the general solution (\ref{EulerSol}), it must satisfy both ultraviolet (UV) and infrared (IR) boundary conditions (\ref{uvirsup}). First, considering the theoretical limit, i.e., $\Lambda=\infty$, we obtain a continuous theory description. In this limit, it can be shown that the general solution satisfies both UV and IR conditions for any $N_f$. However, in actual condensed matter systems, the cutoff is finite and inversely proportional to the lattice constant $a$, where \(a \approx 1 \)\AA. In this case, although the IR condition is still satisfied, the UV condition requires more detailed examination. To further analyze, we redefine the variables $a_{\pm} = -\frac{1}{2} \pm i\frac{R}{2}$, where $R = \sqrt{\frac{N_f^c}{N_f} - 1}$. Depending on the value of $R$, we can distinguish two cases: (i) when $N_f > N_f^c$, $R$ is purely imaginary; (ii) when $N_f \leq N_f^c$, $R$ is real. In the first case, substituting the general solution (\ref{EulerSol}) into the UV condition, we find that only the trivial solution $\sigma(x) = 0$ satisfies the UV condition. Therefore, under a finite cutoff, fermions acquire a nonzero mass, i.e., the system undergoes chiral symmetry breaking, only when $N_f \leq N^c_f$. For two-dimensional Dirac materials like graphene, this is equivalent to opening a gap at the Dirac point, leading to a semimetal-insulator phase transition. For the second case, since $R$ is real, the general solution (\ref{EulerSol}) can be written in the following form:
\begin{equation}
\sigma(x)=\frac{\bar{A}}{\sqrt{x}} \sin \left[R\left(\ln \frac{x}{M_{\Lambda}}+\phi\right)\right].
\end{equation}
where $\bar A=2\sqrt{A_{+}A_{-}}$, $R\phi=\tan^{-1}(-i\frac{A_{+}+A_{-}}{A_{+}-A_{-}})$.
Using the previously redefined $\Sigma(p)\equiv \sigma(pu)/u$, the expression for $\Sigma(p)$ is:
\begin{equation}\label{Sigmaps}
    \Sigma(p) = \frac{\sigma(pu)}{u} = \frac{\bar{A}}{u\sqrt{pu}} \sin \left[R\left(\ln \frac{pu}{M_{\Lambda}}+\phi\right)\right].
\end{equation}
$M_{\Lambda}$ is a parameter with the dimension of mass. Substituting the expression for $\Sigma(p)$ (\ref{Sigmaps}) into the UV boundary condition, it can be found that $M_{\Lambda}$ follows the Miransky scaling law, i.e.:
\begin{equation}
M_{\Lambda}=\Lambda e^{2+R} e^{-2 n \pi / R}.
\end{equation}
where $n$ is some integer. This is analogous to the literature that only considers chiral symmetry breaking induced by PQED interactions (\cite{alves2013chiral}), where numerical results indicate that $\Sigma(0)$ also follows the Miransky scaling law, satisfying $\Sigma(0) \propto \Lambda e^{-1/\sqrt{N_f-N_c}}$. This suggests that as \(N_f \rightarrow N_c\), \(\Sigma(0)\) tends to zero, which also confirms the interpretation of $N_f=N^c_f$ as the critical point for chiral symmetry breaking. Similar results exist for QED\textsubscript{2+1} and QED\textsubscript{3+1} (\cite{maris1994analytic,roberts1994dyson,curtis1990truncating,appelquist1985chiral}).

\begin{figure}[!htbp]
	\centering
	\includegraphics[width=0.4\textwidth]{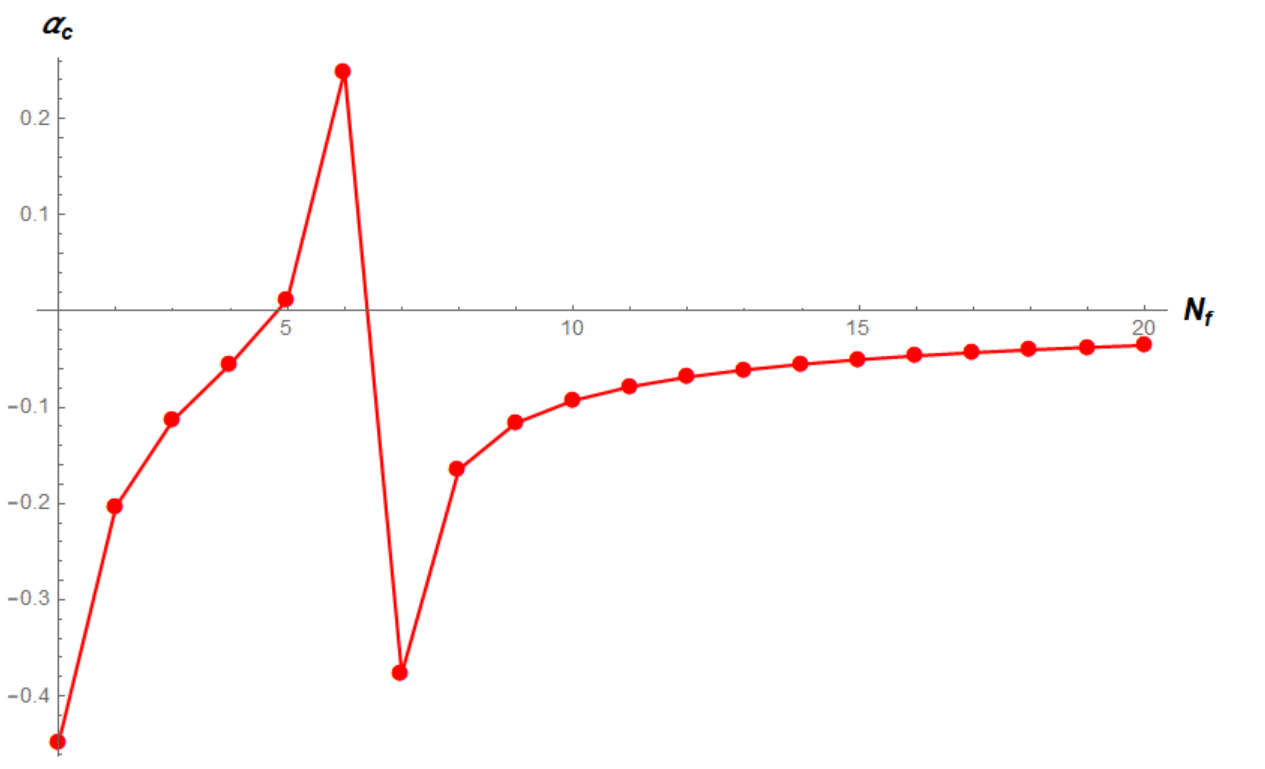}
  \caption{For the case of $N_c = 2$, the graph of the critical fine-structure constant $\alpha_c$ as a function of the fermion flavor number $N_f$.}
\label{alphaNfpng}
\end{figure}

Next, we will explore the crucial role played by non-Abelian $SU(N_c)$ four-fermion interactions in inducing chiral symmetry breaking in electrons. Based on the derived critical fermion number (\ref{Nc}), we obtain the expression for the corresponding critical coupling constant $\lambda_c$ as:
\begin{equation}
\lambda_c=\frac{\pi^2 N_f -32\frac{C(R)}{T(R)}}{2+2D(A)-\frac{\pi^2}{16}N_fD(R)}.
\end{equation}
Considering the substitution $e^2 \rightarrow \lambda/N_f$, the corresponding critical fine structure constant $\alpha_c=\frac{\lambda_c}{4\pi N_f}$ can be expressed as:
\begin{equation}\label{aalpha}
\alpha_c=\frac{1}{4\pi}\frac{\pi^2-32\frac{D(A)}{D(R)N_f}}{2+2D(A)-\frac{\pi^2}{16}N_fD(R)}.
\end{equation}
In the Lagrangian we study (\ref{ssunpqed}), the fermion fields follow the fundamental representation of $SU(N_c)$, hence:
\begin{align}
    D(A)&=N_c^2-1 ,\notag \\
    D(R)&=N_c,\notag \\
    \frac{C(R)}{T(R)}&=\frac{D(A)}{D(R)}=\frac{N_c^2-1}{N_c}.
\end{align}
Therefore, the critical fine structure constant can be expressed as a function of the fermion flavor number $N_f$ and the color number $N_c$ of the non-Abelian gauge group $SU(N_c)$:
\begin{equation}\label{aalpha2}
\alpha_c=\frac{1}{4\pi}\frac{\pi^2-32\frac{N_c ^2-1}{N_c N_f}}{2N_c ^2-\frac{\pi^2}{16}N_f N_c}.
\end{equation}
It is known that chiral symmetry breaking occurs when $\alpha \geq \alpha_c$.
In the pure PQED theory (\cite{alves2013chiral}), chiral symmetry breaking occurs when $\alpha \geq \alpha_c = \pi/4 \approx 0.79$, while incorporating the Gross-Neveu four-fermion interaction reduces this critical value to $\alpha_c \approx 0.36$ (\cite{alves2017dynamical}). Notably, this study finds that the influence of non-Abelian interactions further lowers the value of $\alpha_c$, making chiral symmetry breaking more feasible. We will elucidate this effect through detailed calculations for the $SU(2)$ model. In the case of $SU(2)$, with $N_c=2$, the relationship between $\alpha_c$ and $N_f$ is shown in the figure (\ref{alphaNfpng}).

It can be observed that the maximum value of $\alpha_c$ is $\alpha_c(N_f=6)\approx 0.25$, lower than the aforementioned two critical values, and for $N_f \neq 5,6$, $\alpha_c$ is even less than zero, indicating that non-Abelian interactions greatly promote chiral symmetry breaking in the system, more easily inducing a gap opening in two-dimensional Dirac materials like graphene, leading to a semimetal to insulator phase transition.

Specifically, we also calculated the maximum values of $\alpha_c$ for different $N_c$, as well as the range of critical fermion flavor numbers $N_f$ under different $N_c$. According to formula (\ref{Nc}), we can derive:
\begin{align}
    &N^c _f(\lambda \rightarrow 0)\leq N^c _f(\lambda) \leq  N^c _f(\lambda \rightarrow \infty)\notag \\
    &\Rightarrow 3.24 \frac{C(R)}{T(R)}\leq N^c _f(\lambda)\leq 3.24(\frac{1}{D(R)}+\frac{C(R)}{T(R)})\notag \\
    &\Rightarrow 3.24 \frac{N_c ^2-1}{N_c }\leq N^c _f(\lambda) \leq 3.24 N_c.
\end{align}
It is evident that as the color number $N_c$ of the non-Abelian gauge group increases, the maximum value of $\alpha_c$ correspondingly decreases to below zero, while the minimum value of $N_f ^c$ monotonically increases, further indicating the critical role of non-Abelian interactions in inducing chiral symmetry breaking in two-dimensional Dirac materials like graphene.
The corresponding numerical results are shown in Table (\ref{tab2}) and Table (\ref{tab3s}).

\begin{table}[h]
\centering
\caption{The maximum value of $\alpha_c$ for different $N_c$ values, where the $N_f$ column represents the value that maximizes $\alpha_c$ after fixing $N_c$.}
\label{tab2}
\begin{tabular}{|c|c|c|}
\hline
 & \textbf{$\alpha_{c,max}$} & \textbf{$N_f$} \\ \hline
$N_c=2$ & $0.25$ & $6$ \\ \hline
$N_c=3$ & $0.023$ & $9$ \\ \hline
$N_c=4$ & $-0.0043$ & $12$ \\ \hline
$N_c=10$ & $-0.0009$ & $32$ \\ \hline
\end{tabular}
\end{table}

\begin{table}[h]
\centering
\caption{The maximum and minimum values of $N_f ^c$ for different $N_c$ values.}
\label{tab3s}
\begin{tabular}{|c|c|c|}
\hline
 \textbf{$N_f ^c$}& \textbf{Min} & \textbf{Max} \\ \hline
$N_c=2$ & $4.86$ & $6.28$ \\ \hline
$N_c=3$ & $8.64$ & $9.72$ \\ \hline
$N_c=4$ & $12.15$ & $12.96$ \\ \hline
$N_c=10$ & $32.076$ & $32.4$ \\ \hline
\end{tabular}
\end{table}

\section{\textbf{Summary}}
We have studied the phenomenon of chiral symmetry breaking induced by interactions in two-dimensional Dirac materials, which results in fermions acquiring a dynamically generated mass, thereby opening a gap at the Dirac point and inducing a semimetal-insulator phase transition. We considered two types of interactions, namely PQED electromagnetic interactions and $SU(N_c)$ non-Abelian interactions.

We calculated the propagators of the gauge field and auxiliary boson field using the large $N_f$ expansion and analytically solved the fermion's Schwinger-Dyson self-consistent equation under the rainbow-ladder approximation. This allowed us to obtain an analytical expression for the fermion mass function and derive expressions for the critical fermion flavor number $N^c _f$ and the critical fine structure constant $\alpha_c(N_c,N_f)$ that trigger chiral symmetry breaking.

Our results indicate that, compared to pure PQED and Gross-Neveu four-fermion interactions introduced by impurity effects, $SU(N_c)$ non-Abelian interactions play a significant role in lowering the critical fine structure constant, making the system more susceptible to chiral symmetry breaking. Experimentally, for such non-Abelian interactions, we can simulate them in cold atomic systems (\cite{tagliacozzo2013simulation,banerjee2013atomic,zohar2013cold}), and the opening of the electronic gap due to the semimetal to insulator phase transition can be measured using Bragg spectroscopy.

\section{Acknowledgements}
This work was supported by National Key R\&D Program of China under grants No. 2021YFA1400900, 2021YFA0718300, 2021YFA1402100, NSFC under grants Nos. 61835013, 12174461, 12234012, Space Application System of China Manned Space Program.

\appendix
\section{Derivation of the $SU(N_c)$ Non-Abelian Effective Hamiltonian } \label{aa}
\begin{figure}[H]
    \centering 
    \includegraphics[width=0.43\textwidth]{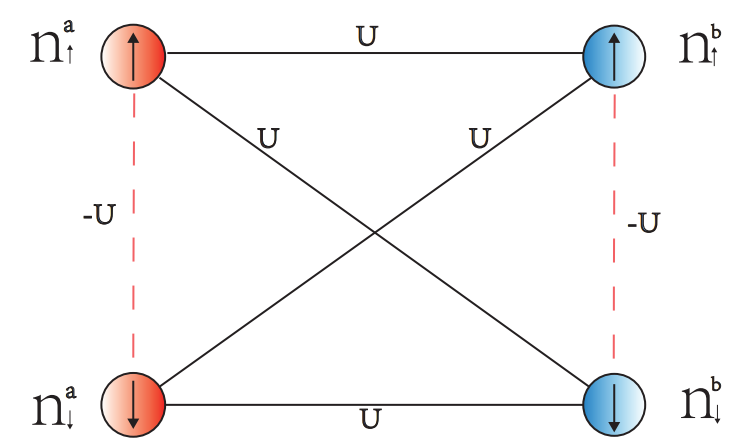} 
    \caption{Consideration of spin degrees of freedom in the Hubbard interaction. The red and blue colors represent the two inequivalent sublattices A and B on the honeycomb lattice, respectively. Solid lines indicate the repulsive Hubbard interactions between different sublattices, while dashed lines denote the attractive Hubbard interactions between different spins on the same sublattice.}
    \label{su22ss}
\end{figure}

\begin{figure}[H]
	\centering
	\includegraphics[width=0.43\textwidth]{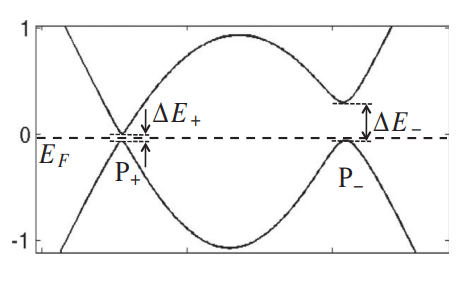}
  \caption{Energy gap at Dirac points modulated by next-nearest-neighbor transitions and magnetic flux. $P_{+}, P_{-}$ represent the two inequivalent Dirac points, with $\Delta E_{+}, \Delta E_{-}$ denoting the corresponding energy gap at $P_{+}, P_{-}$, respectively. Notably, $\Delta E_{-} \gg \Delta E_{+}$, leading to a concentration of the system's low-energy excitations mainly at the $P_{+}$ point. Figure from \cite{palumbo2013abelian}.}
 \label{energes}
 \end{figure}

In reference \cite{palumbo2014non}, it has already been elucidated how to construct the corresponding $SU(N_c)$ Non-Abelian Effective Hamiltonian through the spin-related Hubbard model. Here, we further elaborate on the computational details of the construction.
The Hamiltonian is as follows:
\begin{equation}
    \mathcal{H}=H_0+H_U.
\end{equation}
where $H_0$ is the nearest-neighbor hopping Hamiltonian on the honeycomb lattice:
\begin{equation}
    H_0 = -t \sum\limits_{\left\langle\vec{n}, \vec{n}^{\prime}\right\rangle, \sigma} \left[a_{\vec{n} \sigma}^{\dagger} b_{\vec{n}^{\prime} \sigma} + \text{H.c.} \right].
\end{equation}
The expression for $H_U$ is:
\begin{equation}\label{HUspin}
    H_U=U \left(\sum\limits_{\vec{n},\vec{n}^{\prime},\sigma,\sigma^{\prime}} n^a _{\vec{n},\sigma} n^b _{\vec{n}^{\prime},\sigma^{\prime}} -\sum\limits_{\vec{n},\alpha} n^{\alpha}_{\vec{n},\uparrow}n^{\alpha}_{\vec{n},\downarrow}\right).
\end{equation}
That is, it includes the attractive interaction between different spins on the same sublattice and the repulsive interaction between different sublattices. Its schematic diagram is shown in Fig. (\ref{su22ss}).
The Hamiltonian $H_0$, expanded at the Dirac points, corresponds to the effective Lagrangian:
\begin{equation}
     \mathcal{L}_0=\sum_{\sigma}\bar{\Psi}_{\sigma} \left(i \gamma^{v} \partial_{v}  \right) \Psi_{\sigma}.
\end{equation}
where 
\begin{align}
	\Psi_{\sigma}^T&=(a_{\sigma}\left(K_{+}+\vec{p}\right), b_{\sigma}\left(K_{+}+\vec{p}\right), \notag \\
	&b_{\sigma}\left(K_{-}+\vec{p}\right), a_{\sigma}\left(K_{-}+\vec{p}\right))
\end{align}
 Given that the interaction $H_U$ involves coupling between different spins, for convenience in calculation, we switch to the spin-related spinor: $\Psi_{\alpha}^T=\left(a_{\alpha,\uparrow}, b_{\alpha,\uparrow}, a_{\alpha,\downarrow}, b_{\alpha,\downarrow}\right)$.
Here, $\alpha$ represents the valley degree of freedom, that is, the Dirac points $K,K^\prime$. To simplify the calculations, in the following discussion, we will ignore the $\alpha$ index.
Furthermore, when considering the low-energy characteristics of the system, by employing the large N expansion to handle the interaction, the valley degree of freedom index $\alpha$ can be considered as a type of color degree of freedom. To better match experimental data, we can also modulate the energy gap size at the two inequivalent Dirac points by introducing next-nearest-neighbor transitions and magnetic flux \cite{cirio20143,palumbo2013abelian,palumbo2014non}. Thus, the system's low-energy excitations will be primarily concentrated at one Dirac point, as shown in Fig. (\ref{energes}), thereby allowing the effective SU(2) Thirring model we derive next to more accurately describe the physical properties of the system.

 Next, we will demonstrate that the low-energy effective interaction corresponding to the interaction $H_U$ is of the $SU(2)$ Thirring type non-Abelian interaction. According to equation (\ref{HUspin}), the effective interaction at the Dirac points is:
\begin{equation}\label{HUspinef}
    H_{\text{eff}}=U\sum_{\sigma \sigma^{\prime}} n^{a}_{\sigma}n^{b}_{\sigma^{\prime}}-U\sum_{\alpha}n^{\alpha}_{\uparrow}n^{\alpha}_{\downarrow}.
\end{equation}
Here, for the sake of brevity, we omit the valley degree of freedom index, which can be considered as a degree of freedom for the fermion flavor number.
Next, we will prove that the effective action (\ref{HUspinef}) is actually a non-Abelian interaction, that is, we will show:
\begin{equation}\label{nonabelHU}
    H_{\text{eff}} \equiv \frac{2}{3}U (\bar \psi T^a \gamma^{\mu} \psi)(\bar \psi T^a \gamma^{\mu} \psi).
\end{equation}
Here, $\psi=(\psi_{\uparrow},\psi_{\downarrow})^T=(a_{\uparrow},b_{\uparrow},a_{\downarrow},b_{\downarrow})$, $T^a=\frac{\sigma^a}{2}$ ($a=1,2,3$) are the generators of the $SU(2)$ non-Abelian gauge group, $\gamma^{\mu}$ ($\mu=0,1,2$) are the Dirac matrices, $n^{\alpha}_{\sigma}=\alpha^{\dagger}_{\sigma}\alpha_{\sigma}$ ($\alpha=a,b$) are the particle number operators. In $2+1$ dimensions, we take $\gamma^0=\sigma^3,\gamma^{\nu}=\sigma^{\nu}(\nu=1,2)$.
We will proceed directly from equation (\ref{nonabelHU}) to prove, i.e., to show that equation (\ref{nonabelHU}) is equivalent to equation (\ref{HUspinef}).
By matrix operations, equation (\ref{nonabelHU}) can be expanded as:
\begin{align}
   H_{\text{eff}}&=\frac{2}{3}U \sum_{a} (\bar \psi_i T^a _{ij} \gamma^{\mu} \psi_j)(\bar \psi_l T^a _{lm} \gamma^{\mu} \psi_m) \notag \\
   &=\{(\bar \psi_1 T^1 _{12} \gamma^{\mu} \psi_2+\bar \psi_2 T^1 _{21} \gamma^{\mu} \psi_1)^2+(\bar \psi_1 T^2 _{12} \gamma^{\mu} \psi_2+\bar \psi_2 T^2 _{21} \gamma^{\mu} \psi_1)^2\notag \\
   &+(\bar \psi_1 T^3 _{11} \gamma^{\mu} \psi_1+\bar \psi_2 T^3 _{22} \gamma^{\mu} \psi_2)^2\}.
\end{align}
Substituting $T^a=\frac{\sigma^a}{2}$ into the above equation yields:
\begin{align}\label{u6s}
	H_{\text{eff}}&=\frac{U}{6}\{4(\bar \psi_1 \gamma^{\mu}\psi_2)(\bar \psi_2 \gamma^{\mu}\psi_1)+(\bar \psi_1 \gamma^{\mu}\psi_1)^2 \notag \\
	&+(\bar \psi_2 \gamma^{\mu}\psi_2)^2-2(\bar \psi_1 \gamma^{\mu}\psi_1)(\bar \psi_2 \gamma^{\mu}\psi_2)\}
\end{align}
Using $\psi_1=\psi_{\uparrow}=(a_{\uparrow},b_{\uparrow})$ and $\psi_2=\psi_{\downarrow}=(a_{\downarrow},b_{\downarrow})$ along with the anticommutation relations between fermion operators:
\begin{align}
	\{a_{\sigma},a^{\dagger}_{\sigma^{\prime}}\}&=\delta_{\sigma\sigma^{\prime}},\notag \\
		\{a_{\sigma},b_{\sigma^{\prime}}\}&=0, \notag \\
		 \{a_{\sigma},a_{\sigma^{\prime}}\}&=0, \notag \\
		\{a_{\sigma},b^{\dagger}_{\sigma^{\prime}}\}&=0.
	\end{align}
Equation (\ref{u6s}) can be expanded as follows:
\begin{align}
    H_{\text{eff}}&=\frac{U}{6}\{ 4(2a^{\dagger}_{\uparrow} a_{
    \uparrow}b^{\dagger}_{\downarrow}b_{
    \downarrow}+2a^{\dagger}_{\downarrow} a_{
    \downarrow}b^{\dagger}_{\uparrow}b_{
    \uparrow}-b^{\dagger}_{\uparrow}b_{
    \uparrow}b^{\dagger}_{\downarrow}b_{
    \downarrow}-a^{\dagger}_{\uparrow}a_{
    \uparrow}a^{\dagger}_{\downarrow}a_{
    \downarrow}\notag \\
	&+a^{\dagger}_{\downarrow}a_{
    \uparrow}b^{\dagger}_{\uparrow}b_{
    \downarrow}+a^{\dagger}_{\uparrow}a_{
    \downarrow}b^{\dagger}_{\downarrow}b_{
    \uparrow})\notag \\
    &+(6a^{\dagger}_{\uparrow} a_{
    \uparrow}b^{\dagger}_{\uparrow}b_{
    \uparrow}+6a^{\dagger}_{\downarrow} a_{
    \downarrow}b^{\dagger}_{\downarrow}b_{
    \downarrow})\notag \\
    &-4a^{\dagger}_{\downarrow}a_{
    \uparrow}b^{\dagger}_{\uparrow}b_{
    \downarrow}-4a^{\dagger}_{\uparrow}a_{
    \downarrow}b^{\dagger}_{\downarrow}b_{
    \uparrow}-2b^{\dagger}_{\uparrow}b_{
    \uparrow}b^{\dagger}_{\downarrow}b_{
    \downarrow}\notag \\
	&-2a^{\dagger}_{\uparrow}a_{
    \uparrow}a^{\dagger}_{\downarrow}a_{
    \downarrow}-2a^{\dagger}_{\uparrow} a_{
    \uparrow}b^{\dagger}_{\downarrow}b_{
    \downarrow}-2a^{\dagger}_{\downarrow} a_{
    \downarrow}b^{\dagger}_{\uparrow}b_{
    \uparrow}\notag \\
&=U(n^{a}_{\uparrow}n^{b}_{\uparrow}+n^{a}_{\uparrow}n^{b}_{\downarrow}+n^{a}_{\downarrow}n^{b}_{\uparrow}+n^{a}_{\downarrow}n^{b}_{\downarrow}-n^{a}_{\uparrow}n^{a}_{\downarrow}-n^{b}_{\uparrow}n^{b}_{\downarrow}) \notag \\
    &\equiv U\sum_{\sigma \sigma^{\prime}} n^{a}_{\sigma}n^{b}_{\sigma^{\prime}}-U\sum_{\alpha}n^{\alpha}_{\uparrow}n^{\alpha}_{\downarrow}.
\end{align}
At this point, we have demonstrated that the effective interaction, which accounts for the attractive interaction between different spins on the same sublattice and the repulsive interaction between different sublattices (\ref{HUspin}), is non-Abelian. It is described by the $SU(2)$ Thirring model (see equation (\ref{nonabelHU})).

\bibliographystyle{apsrev4-1}
\bibliography{refer}
	
\end{document}